BMC
Plant Biology

# Transcriptomic and metabolomic analysis of copper stress acclimation in *Ectocarpus siliculosus* highlights signaling and tolerance mechanisms in brown algae

Andrés Ritter[1,2,3,5], Simon M Dittami[1,2], Sophie Goulitquer[4], Juan A Correa[3], Catherine Boyen[1,2], Philippe Potin[1,2] and Thierry Tonon[1,2*]

## Abstract

**Background:** Brown algae are sessile macro-organisms of great ecological relevance in coastal ecosystems. They evolved independently from land plants and other multicellular lineages, and therefore hold several original ontogenic and metabolic features. Most brown algae grow along the coastal zone where they face frequent environmental changes, including exposure to toxic levels of heavy metals such as copper (Cu).

**Results:** We carried out large-scale transcriptomic and metabolomic analyses to decipher the short-term acclimation of the brown algal model *E. siliculosus* to Cu stress, and compared these data to results known for other abiotic stressors. This comparison demonstrates that Cu induces oxidative stress in *E. siliculosus* as illustrated by the transcriptomic overlap between Cu and $H_2O_2$ treatments. The common response to Cu and $H_2O_2$ consisted in the activation of the oxylipin and the repression of inositol signaling pathways, together with the regulation of genes coding for several transcription-associated proteins. Concomitantly, Cu stress specifically activated a set of genes coding for orthologs of ABC transporters, a $P_{1B}$-type ATPase, ROS detoxification systems such as a vanadium-dependent bromoperoxidase, and induced an increase of free fatty acid contents. Finally we observed, as a common abiotic stress mechanism, the activation of autophagic processes on one hand and the repression of genes involved in nitrogen assimilation on the other hand.

**Conclusions:** Comparisons with data from green plants indicate that some processes involved in Cu and oxidative stress response are conserved across these two distant lineages. At the same time the high number of yet uncharacterized brown alga-specific genes induced in response to copper stress underlines the potential to discover new components and molecular interactions unique to these organisms. Of particular interest for future research is the potential cross-talk between reactive oxygen species (ROS)-, *myo*-inositol-, and oxylipin signaling.

* Correspondence: tonon@sb-roscoff.fr
[1]UPMC Univ Paris 06, UMR 8227, Integrative Biology of Marine Models, Station Biologique de Roscoff, Sorbonne Universités, CS 90074, F-29688 Roscoff cedex, France
[2]CNRS, UMR 8227, Integrative Biology of Marine Models, Station Biologique de Roscoff, CS 90074, F-29688 Roscoff cedex, France
Full list of author information is available at the end of the article

## Background

Brown algae (Kingdom Chromalveolata, Division Heterokontophyta, Class Phaeophyceae) are sessile macro-organisms of great ecological relevance in coastal ecosystems, and belong to an evolutionary lineage that has diverged from land plants and other multicellular organisms more than one billion years ago through secondary endosymbiosis [1]. Most of these seaweeds grow in the intertidal zone where they must face constant abiotic fluctuations, e.g. in temperature, irradiation, and salinity, in relation with the tidal cycle. In addition to these natural constraints, they must deal with pollutants, including heavy metals (HMs), resulting from human activities. These pollutants represent a major threat for marine ecosystems, impacting benthic flora and fauna assemblages [2-4]. Copper (Cu) is a vital micronutrient, essential for all forms of life. It acts as cofactor for many enzymatic systems, and participates in crucial physiological processes including photosynthesis and respiration. However, excessive Cu concentrations are harmful for most living organisms. For this reason, free Cu is found only in traces in eukaryotic cells, where it is tightly controlled by a set of specific transporters and cytosolic chaperones that deliver it to their respective target proteins or organelles [5]. At high concentrations, both cupric and cuprous (Cu(II)/Cu(I), $E^0 = +0.15$ V) ions can participate in redox reactions affecting organisms at the cellular level mainly by three well-established processes: (i) direct protein inactivation by undesired amino acid-metal interactions due to the affinity of Cu(II) for thiol-, imidazole-, and carboxyl- groups; (ii) in presence of superoxide or reducing agents Cu(II) can be reduced to Cu(I), which is capable of catalyzing the formation of hydroxyl radicals via the non-enzymatic Fenton's reaction [6]; (iii) displacement of essential cations from specific binding sites. These effects impact a wide range of cellular processes in photosynthetic organisms, interfering for instance with fatty acid and protein metabolism, or inhibiting respiration and nitrogen fixation processes [7]. Copper affects in particular photosystem electron transfer components, leading to the generation of reactive oxygen species (ROS) and peroxidation chain reactions involving membrane lipids [8,9]. For these reasons, this metal has been extensively utilized as antifouling agent to prevent the proliferation of algal flora on immersed surfaces.

Most organisms deploy an array of mechanisms to control cellular Cu levels, for detoxification, and to repair damaged cellular structures, which are triggered *via* the activation of signaling pathways. Signal transduction involves elements shared by plants and animals, but also molecules specific to each lineage [10]. Plants may trap free Cu by increasing levels of chelating agents such as chaperones, metallothioneins, phytochelatins, or organic acids [11]. This process is linked to an increase of transmembrane activity, in which Cu $P_{1B}$-type ATPases and multidrug resistance ABCC transporters (formerly known as multidrug resistance-related proteins) may have key roles to sequester or exclude chelated forms of Cu or other toxic adducts [12,13].

Knowledge of the molecular bases of Cu stress regulation in brown algae is still scarce and scattered. Previous studies in this domain have focused on a few specific physiological aspects such as photosynthesis [14,15], oxidative stress [16,17], or metal chelation [18,19], and most constitute field studies of long-term adaptation to chronic Cu exposure. Recently, two publications have reported large-scale proteomic analyses to identify mechanisms underlying acclimation to high Cu levels [20,21]. These studies showed the increase of brown algal specific antioxidant mechanisms and changes in photosynthesis-related proteins to cope with chronic Cu stress. However, there is still limited molecular data, especially at the transcriptomic level, on how brown algae sense short-term variations in metal content and induce the regulation of intracellular Cu concentrations through specific signaling processes. Previous work on the brown alga *Laminaria digitata* showed that short term exposure to Cu stress triggers the synthesis of 12-oxophytodienoic acid (12-OPDA) and prostaglandins, concomitantly with changes in expression of selected genes involved in stress response [22]. These results suggest that brown algae may synthesize plant-like octadecanoid, but also eicosanoid oxylipins, to induce stress-related detoxification responses.

So far, global molecular analyses of the brown algal stress response were hampered by the lack of genomic resources. The development of the biological model *Ectocarpus siliculosus*, including access to its genome sequence, represents a major breakthrough in algal research, and opens the gates to "omics"-based approaches [23,24]. Regarding Cu-homeostasis, the annotation of the *E. siliculosus* genome allowed identifying several putative Cu-chaperones, ABC transporters, and Cu-channels controlling cellular Cu traffic. Moreover, two recent reports by Dittami et al. [25,26] have established a starting point for the integrated analysis of transcript and metabolite profiling during the short-term response to saline and oxidative stress in *E. siliculosus*. They demonstrated the repression of primary metabolism and the activation of recycling of existing proteins through autophagy under the abiotic stress conditions tested.

In the present study, we employed large-scale transcriptome and metabolome analyses to gain insights into the short-term acclimation of *E. siliculosus* to Cu stress, and extended the analysis of gene expression data to results previously published for other abiotic stressors. The integration of these results highlighted the Cu induction of a large panel of signaling mechanisms likely to constitute the driving force behind the observed transcriptomic and metabolic shifts. In particular, oxylipin metabolism and several distinct genes coding for transcription-associated proteins were up-regulated. Moreover, transcriptomic meta-analysis of Cu and other abiotic stressors showed tight links between Cu and oxidative stress, and confirmed previous observations such as the repression of genes encoding enzymes involved in primary amino acid biosynthesis, balanced with the induction of autophagic processes. In addition, this analysis allowed the identification of Cu stress specific mechanisms such as the up-regulation of multidrug resistance ABC transporters, putative Cu $P_{1B}$-type ATPases, and of a vanadium bromoperoxidase involved in halide metabolism.

## Results

### Cu stress treatments alter photosynthetic capabilities of *E. siliculosus*

To monitor the short-term acclimation to stress rather than cell death, the intensity of Cu stress had to be carefully selected. In previous chronic Cu stress experiments, we showed that Cu(II) at a concentration of $250\,\mu g\,L^{-1}$ led to a drastic decrease of photosynthetic activity (Fv/Fm) after 1 to 6 days of treatment [21]. In this study, we aimed to monitor the acute Cu stress response of *E. siliculosus*. Therefore algae were incubated in presence of Cu(II) during 8 h at final concentrations of $250\,\mu g\,L^{-1}$ and at $500\,\mu g\,L^{-1}$. These concentrations are comparable to those registered in Cu polluted marine sites where *E. siliculosus* has been observed [27,28], and correspond to an approximately 400-fold enrichment of the total dissolved Cu content in the natural seawater used to set-up the experiments. Changes in the photosynthetic yield were then followed on an hourly basis. Compared to the control treatment, algae incubated with $500\,\mu g\,L^{-1}$ displayed significantly lower Fv/Fm ratios after 6 h (U test, $p < 0.05$), while $250\,\mu g\,L^{-1}$ induced a significant decrease (U test, $p < 0.05$) only after 8 h of treatment (Figure 1). Taking these results into consideration, algae were sampled after 4 h and 8 h of incubation in medium with $250\,\mu g\,L^{-1}$ Cu(II) to monitor molecular changes occurring prior and during the observed reduction of photosynthetic activity.

### Meta-transcriptomic analysis highlights copper specific genes, shared responses with oxidative stress treatments, and core abiotic stress genes

Copper stress treatment resulted in 909 significantly regulated contigs/singletons under Cu stress (regardless of time), 546 of which were up-regulated and 363 down-regulated compared to control conditions (two-way ANOVA, FDR of 5%). None of the examined contigs/singletons showed a significant interaction between time and treatment, indicating little difference in the transcriptomic response to 4 h and 8 h of Cu treatment. Only 11% of the contigs/singletons induced in response to Cu stress were automatically classified by the GOLEM and KOBAS annotation tools (data not shown); therefore manual classification was done for genes presenting a fold-change ratio > 2 compared to the control conditions. This dataset represents 560 genes (627 contigs/singletons), which were assigned to 13 categories (Figure 2, Additional file 1).

Cu stress transcriptomic data were further compared to previous results obtained with the same array and protocol for short-term (6 h) hyposaline (56 mM NaCl final concentration), hypersaline (1,470 mM NaCl final concentration) and oxidative (1 mM $H_2O_2$ final concentration) stress [25] [ArrayExpress:E-TABM-578]. To reduce the number of false negatives no FDR correction was applied for this meta-analysis, and all transcripts previously found to be significantly regulated (p < 0.05, fold-change compared to control > 2) after either 6 h [25] or 4 or 8 h (this study) were taken into account. Seventy-five contigs/singletons were up-regulated in all considered stress conditions (Additional file 2A). Seventy-six percent of them (i.e. 57) code for proteins with unknown functions, and 48% (i.e. 36) are orphans (Additional file 3), that is as of March 2013, they had no match outside *E. siliculosus* in the NCBI nr database at an e-value cutoff of 1e-5. Regarding down-regulated transcripts, we identified 64 contigs/singletons (Additional file 2B) of which 55% (i.e. 35) correspond to unknown proteins and 27% (i.e. 17) are orphans.

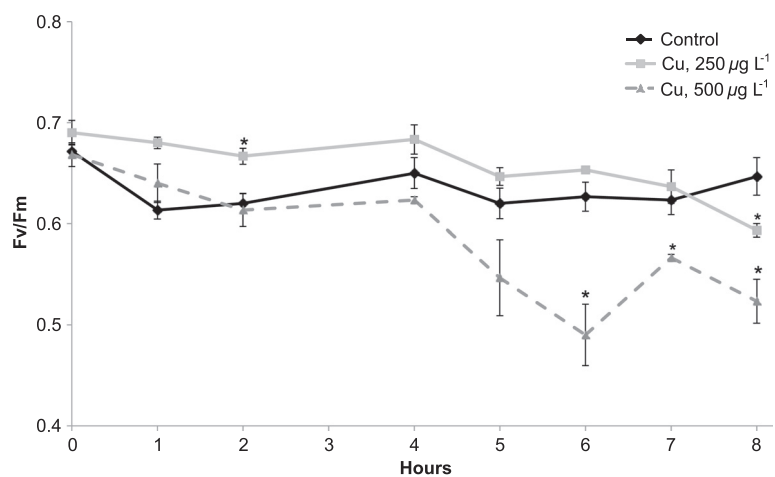

**Figure 1 Physiological effects of acute copper (Cu) toxicity on *E. siliculosus* strain Ec32.** Changes in the photosynthetic yield (Fv/Fm) were monitored during 8 h in absence of Cu (diamonds), and in presence of 250 (squares) and 500 $\mu g\ L^{-1}$ (triangles) of CuCl$_2$ (final concentration). Values represent means of three independent replicates and bars represent the standard error. Asterisks highlight significantly different values from the respective control condition (U-test, $p < 0.05$).

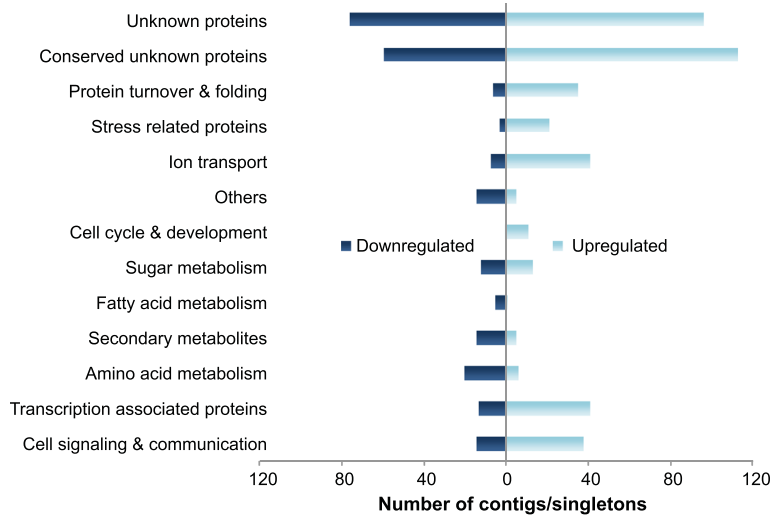

**Figure 2 Functional distribution of contigs/singletons up-regulated and down-regulated in Cu-stressed *E. siliculosus*.** Contigs/singletons were manually classified according to their annotation in the *E. siliculosus* genome database. The figure shows the number of significantly down- (left) and up-regulated (right) contig/singletons in each category (two-way ANOVA, FDR < 5%).

Hierarchical clustering (Additional file 4) revealed that $H_2O_2$ stress samples formed a cluster with Cu-stress samples, likely reflecting the fact that Cu(II) treatment unbalances the cellular redox state, leading to oxidative stress. Indeed, 42% of the contigs/singletons regulated (336 of the induced and 314 of the repressed) in response to copper stress were also regulated at least in response to $H_2O_2$ (Additional file 1). In the following sections, we focus on the manual analysis of selected gene families or functional categories of genes representing three profiles of expression: genes regulated by all available stressors, genes regulated at least by Cu and $H_2O_2$, and finally genes specifically regulated by copper stress. Lists of these genes as well as available annotations are provided in Additional file 3. Gene set enrichment analyses were performed for these three different groups of genes using Blast2GO, but did not yield any significantly overrepresented GO terms at an FDR < 5%.

## Cell signaling and communication

A phospholipase C (PLC) gene (comprising two genomic loci, Esi0000_0131 and Esi0000_0133) was up-regulated specifically under Cu stress. The locus Esi0085_0013, encoding a calcium-dependent phospholipid binding protein of the Copine family that is related to abiotic stress processes in land plants, was induced under both Cu and $H_2O_2$ stress. In addition, a gene encoding a Ca-dependent protein kinase (Esi0073_0115) and a serine threonine kinase (Esi0038_0037) were up-regulated by both stressors. Conversely, a large set of genes involved in *myo*-inositol metabolism were down-regulated under these treatments. This was true for a cytidine diphosphate diacylglycerol (CDP-DAG) synthase (Esi0016_0173), a putative phosphatidylinositol-4-phosphate (PI4P) 5-kinase (Esi0030_0108), a *myo*-inositol-1-phosphate synthase (Esi0279_0022), a phosphatidylinositol transfer protein SEC14 (Esi0499_0013), and a putative inositol monophosphatase (Esi0133_0061).

In relation to downstream lipid metabolism and signaling cascades, a gene encoding a cytochrome P450 enzyme (Esi0060_0078) was among the most highly up-regulated genes during copper treatment (41-fold after 4 h, 8-fold after 8 h) and also responded to oxidative stress (Additional file 3). The corresponding protein displays approximately 30% identity and 45% similarity with the cytochrome P450 domain of plant allene oxide synthases (AOSs) of the CYP74 family involved in the jasmonate biosynthetic pathway. A multiple alignment of this protein with members of the CYP74 family highlights the existence of several motifs common to AOSs, such as the IHCD motif and the conserved F and S catalytic amino acid residues (Additional file 5A). Finally, protein structure homology modelling using the Phyre2 software predicts high structural similarity between Esi0060_0078 and the AOS of *Parthenium argentatum* (Additional file 5, panel B; PDB: 3dan; 100% confidence, 91% coverage).

### Transcription factors (TFs)

Among the 283 TFs in the *E. siliculosus* genome [29], our study identified 19 up-regulated and 6 down-regulated genes (Table 1). Eight of these genes were specifically up-regulated by Cu stress, whereas seven others were co-induced by Cu and $H_2O_2$. Among them, two potential arsenite-inducible AN1-ZFPs, Esi0002_0015

(specifically up-regulated by Cu), and Esi0348_0030 (induced by Cu, $H_2O_2$ and hyposaline stress) were found. Interestingly, Esi0002_0015 was more similar to animal AN1-ZFPs, and Esi0348_0030 to plant-type stress associated proteins. Also in relation to stress TFs, the heat shock factor Esi0279_0021 was among the genes most strongly repressed in response to Cu stress, with a 26-fold change after 4 h of treatment (Table 1). Furthermore, two Cu specifically up-regulated genes, Esi0100_0085 and Esi0226_0031, contained the RWP-RK domain found in plant proteins involved in nitrogen-controlled development [30]. Finally, we recorded the transcriptional up-regulation of a gene coding for a protein containing a zinc finger

**Table 1 Significantly (FDR < 0.05) Cu-regulated transcription factors (TFs) in *E. siliculosus***

| Gene ID | TF family | Domain 1 | Domain 2 | Annotation | Log2-ratio 4 h | Log2-ratio 8 h | Meta-analysis |
|---|---|---|---|---|---|---|---|
| Esi0219_0040 | bZIP | bZIP_1 | bZIP_2 | Conserved unknown protein | 1.1 | 0.6 | Cu_Oxi_up |
| Esi0051_0089 | C2H2 | zf-C2H2 | - | Conserved unknown protein | 1.2 | 0.3 | Cu_up |
| Esi0348_0030 | C2H2 | zf-AN1 | zfAN1 | Arsenite inducible RNA associated protein AIP-1-related SAP | 1.1 | 0.7 | Cu_Hypo_Oxi_up |
| Esi0002_0015 | C2H2 | zf-AN1 | - | Arsenite inducible RNA associated protein AIP-1-related SAP | 1.8 | 1.8 | Cu_up |
| Esi0513_0010 | C2H2 | zf-C2H2 | - | Similar to metal response element-binding transcription factor-1 MTF1 | 1.4 | 0.8 | Cu_Oxi_up |
| Esi0301_0006 | C2H2 | R3H | - | Conserved unknown protein | 2.0 | 1.4 | Cu_Oxi_up |
| Esi0040_0071 | C2H2 | C2H2 | - | Expressed unknown protein | 1.8 | 1.8 | Cu_Hyper_oxi_up |
| Esi0071_0018 | C2H2 | C2H2 | PUB | PUB domain, zinc finger protein thioredoxin | 1.5 | 1.3 | Cu_Hypo_up |
| Esi0100_0023 | CCHC | CCHC | - | Conserved unknown protein | 1.3 | −0.1 | Cu_Hypo_oxi_up |
| Esi0151_0061 | CCHC | CCHC | - | Conserved unknown protein | 1.4 | 0.7 | Cu_Hypo_up |
| Esi0118_0050 | C3H | zf-CCCH | - | Conserved unknown protein | 1.1 | 0.3 | Cu_up |
| Esi0201_0038 | CCAAT_HAP3 | NF-YB | - | Conserved unknown protein | 1.8 | 1.4 | Cu_Hypo_up |
| Esi0292_0018 | FHA | FHA | - | Nibrin | 1.3 | 0.6 | Cu_up |
| Esi0041_0031 | GNAT | Acetyltransf_1 | - | Conserved unknown | 1.3 | 1.7 | Cu_up |
| Esi0226_0031 | RWP-RK | RWP-RK | - | NIN-like transcription factor | 1.1 | 0.3 | Cu_up |
| Esi0100_0085 | RWP-RK | RWP-RK | - | NIN-like transcription factor 4 | 1.4 | 0.5 | Cu_up |
| Esi0004_0202 | SWI/SNF_SNF2 | Helicase_C | SNF2_N | DEAD-like helicase | 0.9 | 1.1 | Cu_up |
| Esi0071_0075 | Zn_clus | Zn_clus | - | Conserved unknown | 3.1 | 1.6 | Cu_Oxi_up |
| Esi0071_0081 | Zn_clus | Zn_clus | - | Conserved unknown | 1.4 | 0.4 | Cu_up |
| Esi0279_0021 | HSF | HSF | - | Heat Shock transcription factor | −4.7 | −2.0 | Cu_down |
| Esi0149_0075 | JmjC | JmjC | - | Transcription factor jumonji/ aspartyl beta-hydroxylase | −1.3 | −0.9 | Cu_down |
| Esi0290_0010 | mTERF | mTERF | - | Conserved unknown protein | −2.7 | −0.7 | Cu_Hypo_Oxi_down |
| Esi0095_0057 | MYB-related | Myb | - | Conserved unknown protein | −2.4 | −1.3 | Cu_Hypo_Oxi_down |
| Esi0013_0140 | RWP-RK | RWP-RK | - | NIN-like 6 | −2.6 | −1.6 | Cu_Hypo_Oxi_down |
| Esi0356_0029 | AP2/EREBP | AP2 | - | Pathogenesis-related transcriptional factor and ERF | −1.9 | −1.1 | All_down |

Log2-ratios represent the means of three biological replicates. In the column "Meta-analysis", Cu is for copper stress, Oxi for oxidative stress, Hyper for hypersaline stress, and Hypo for hyposaline stress condition. The direction of the changes under these different treatments is indicated by up and down for induction or repression, respectively.

domain that was initially annotated as a metal responsive transcription factor (MTF)-1-like protein (Esi0513_0010) under both Cu and oxidative stress. However, although this protein showed some similarities to the DNA binding domain of metazoan MTF-1, no further homology to MTF-1 proteins was observed outside this domain, making it impossible to infer the function of Esi0513_0010 without further experimentation.

### Stress and detoxification mechanisms

Many genes coding for proteins involved in general stress and detoxification mechanisms were regulated under both Cu and $H_2O_2$ stress. For instance, eight heat shock proteins (HSPs) of the 70, 40, and 20 classes were up-regulated, possibly to facilitate the refolding of damaged proteins and to prevent protein aggregation (Additional file 3) under acute oxidative stress. The up-regulation of a DNA double-strand break repair rad50 ATPase (Esi0002_0198) was also observed. Finally, genes coding for proteins with important functions for antioxidant mechanisms were also induced by Cu and $H_2O_2$, including a glutathione reductase (Esi0019_0176), two glutathione-S-transferases (Esi0648_0004 and Esi0191_0054), and one thioredoxin (Esi0030_0031).

Besides the overlapping responses to $H_2O_2$ and Cu treatments, several genes regulated exclusively by Cu stress were identified, including two chloroplastic iron-dependant superoxide dismutases (Fe-SODs; Esi0219_002 and Esi0201_0013) and a vanadium dependent bromoperoxidase (vBPO; Esi0009_0080) which is a ROS detoxifying enzyme specific of brown algae (Additional file 3). Furthermore, a glutathione-S-transferase (Esi0002_0065), and two glutaredoxins, Esi0050_0061 (a glutaredoxin/malate transporter fusion protein) and Esi0036_0002, were also specifically up-regulated by Cu. These latter enzymes are likely involved in HM-detoxification functions such as reduction of Cu-glutathione adducts and export of Cu-organic acid adducts. Several additional stress-related transport systems were specifically up-regulated by Cu stress. Among them were a multidrug and toxic compound extrusion protein (MATE; Esi0017_0140) and a putative heavy metal $P_{1B}$-type ATPase (HMA; Esi0023_0054) (Additional file 3). This protein contains a conserved HMA motif in the N-terminal cytoplasmic region, a central E1-E2 ATPase domain, and a haloacid dehalogenase domain in the C-terminal region (Additional file 6); PSI-BLAST shows 41% of identity between the *E. siliculosus* protein and the *A. thaliana* $Cu^+$ exporting $P_{1B}$ type ATPase HMA5 (AT1G63440).

We also observed Cu specific up-regulation of five of the 69 putative *E. siliculosus* ABC transporter proteins (ABCTs) (Additional file 3). Phylogenetic analysis, together with reference sequences from the human [31] and *A. thaliana* [32] superfamily of ABCTs, led to the classification of these five *E. siliculosus* proteins into four ABCT subfamilies (Figure 3). Of particular interest are Esi0109_0024 and Esi077_0044, which fall into the stress-detoxification ABC-B (MDR/TAP) and ABC-C (CFTR/MRP) subfamilies, respectively. In addition, Esi0359_0018 and Esi0154_0007 clustered with plant and human ABC-A transporters.

### Primary metabolism and protein turnover

As illustrated by our PAM measurements, one basic process altered by copper stress was photosynthesis, which is directly linked to the generation of ROS when intracellular copper concentrations are not properly regulated. In our study, two genes coding for stress-related chlorophyll binding proteins (CBPs) of the LI818 family, Esi0002_0349 and Esi0085_0016, appeared up-regulated specifically by Cu stress, and a third, Esi0085_0049, was up-regulated under copper, oxidative, and hyposaline stress. A fourth Cu-induced CBP, Esi0458_0016, belonged to the LHCF clade [33]. We also observed a trend for down-regulation, at least under Cu stress, of a number of genes related to the xanthophyll cycle such as putative violaxanthin de-epoxidases, as well as a putative zeaxanthin epoxidase (Additional file 1). Regarding chlorophyll biosynthesis, Cu and $H_2O_2$ repressed genes encoding an Mg chelatase, a chlorophyll synthase, and a protoporphyrinogen oxidase. Although these changes did not pass our rather strict FDR correction, they were significant when analysed individually (Additional file 3).

Twenty of the genes repressed by Cu stress were associated to nitrogen assimilation and primary amino acid synthesis (Figure 4). Several of them were also down-regulated under $H_2O_2$ stress, such as one nitrite reductase (NAD(P)H and ferredoxin; Esi0249_0028), one glutamate synthase (Esi0029_0131), one glutamate dehydrogenase 1 (Esi0028_0164), and three putative $NH_4^+$ and $NO_3^-$ transporters (Esi0526_0006, Esi0278_0026, and Esi0278_0032) (Additional file 3). These observations suggest the repression of the GS/GOGAT pathway under oxidative stress conditions. In addition, an agmatinase (Esi0039_0062), and a spermine synthase (Esi0000_0445) were specifically down-regulated by Cu, indicating the repression of the arginine/ornithine-derived polyamine pathway. Three other genes, coding for proteins involved in nitrate, ammonium, urea and amino acid transport (Esi0278_0026, Esi0526_0006, and Esi0104_0047), were down-regulated in all stress conditions included in our meta-analysis.

In contrast to the transcriptomic repression of nitrogen assimilation, 35 contigs/singletons related to protein turnover processes (Figure 2) were induced in response to copper stress. Among them, 12 genes correspond to the ubiquitin system, and all of them except one encoding an ubiquitin related protein Esi0009_0093 were induced in response to several stressors (Additional file 3).

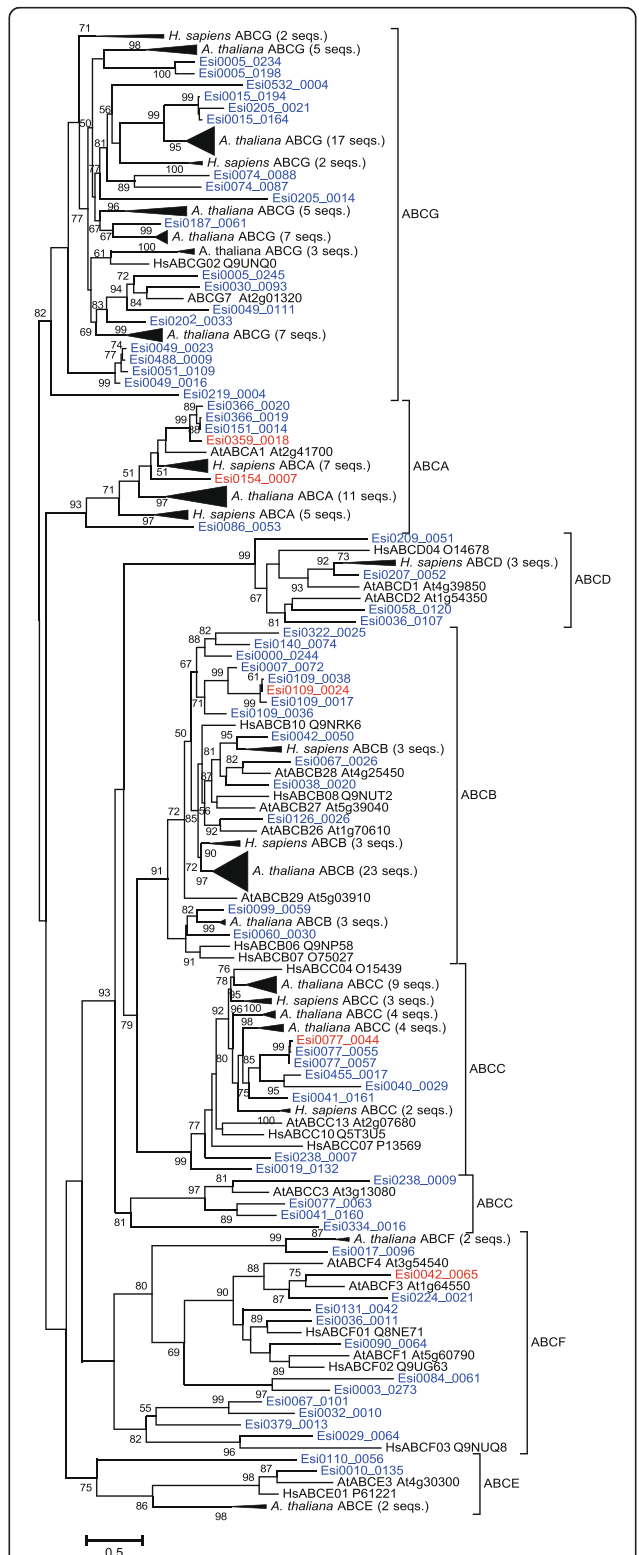

**Figure 3 Phylogenetic tree of human, *A. thaliana*, and *E. siliculosus* (red and blue) ABC transporters.** *E. siliculosus* genes induced specifically by Cu stress are marked in red. Confidence values are the results of an approximate likelihood ratio test; only confidence values ≥ 50 are shown.

Eight genes corresponded to proteasome-related proteins, suggesting the induction of stress-related autophagy mechanisms. Four of these genes were induced specifically in response to copper. Furthermore, we detected the up-regulation of six RING type domain proteins that might correspond to E3-ligases (Additional file 1).

Parallel to changes in N metabolism, we observed down-regulation of genes involved in fatty acid synthesis and degradation. Moreover, twenty-one genes related to sugar metabolism were significantly regulated in response to Cu treatments (FDR < 5%; Additional file 1). Most of them correspond to proteins involved in central carbon metabolism and storage, sugar transport, and structural modifications of cell wall polysaccharides, and were also regulated in response to other stressors (Additional file 3). Regarding sugar metabolism, it is interesting to note the presence of genes involved in mannitol and trehalose synthesis, both repressed in presence of Cu (Additional file 1). With respect to cell wall structure modifications, we noted that Cu and $H_2O_2$ treatments induced the up-regulation of two genes encoding the alginate modifying enzymes mannuronan-C5-epimerases MEP6 and MEP7 (Additional file 3). A number of loci corresponding to other polysaccharide modifying enzymes were also induced in presence of copper, mainly after 4 h of treatment, such as several loci encoding glycoside hydrolases (GHs) and glycoside transferases (GTs) (Additional file 3). However, none of these Cu-regulated GTs or GHs was induced in presence of $H_2O_2$.

## Cu-induced changes at the transcriptome level modulate metabolite composition

To relate the transcriptional reprogramming induced by copper stress in *E. siliculosus* with changes in the metabolome, metabolite profiling was carried out for algal samples harvested after 4 h and 8 h of treatments. UPLC-MS experiments in positive ion mode for Cu-stressed algae provided 392 monoisotopic peaks. Partial least squares discriminant analysis (Additional file 7) and hierarchical clustering (Additional file 8) of these compounds indicated a clear distinction between Cu stress and control samples, but the different treatment times could not be separated. Two major groups of metabolites emerged from this analysis, corresponding to compounds that either accumulate or decline in response to Cu stress. Within each of these groups two smaller clusters, with either a strong or a weak response to stress, were visible. The combination of UPLC-MS and GC-MS analysis allowed us to identify 47 compounds corresponding to fatty acids, oxylipins, and amino acids (Figure 5).

## Changes in amino acid contents

In agreement with our results obtained for the global metabolite profiles, PLS-DA of samples according to their amino acid profile allowed to discriminate between control and stress conditions (Additional file 9, panel A), as well as between durations of treatments. The primary amino acids in *E. siliculosus*, alanine and glutamate [34] showed no significant variations in response to stress (p > 0.05; Figure 5). In contrast, contents of aromatic (phenylalanine and tyrosine) and branched chain amino acids (valine, leucine, and isoleucine) increased at least 2 fold after 4 h and 8 h of Cu treatment (two-way ANOVA, p < 0.05), likely as a result of increased protein catabolism.

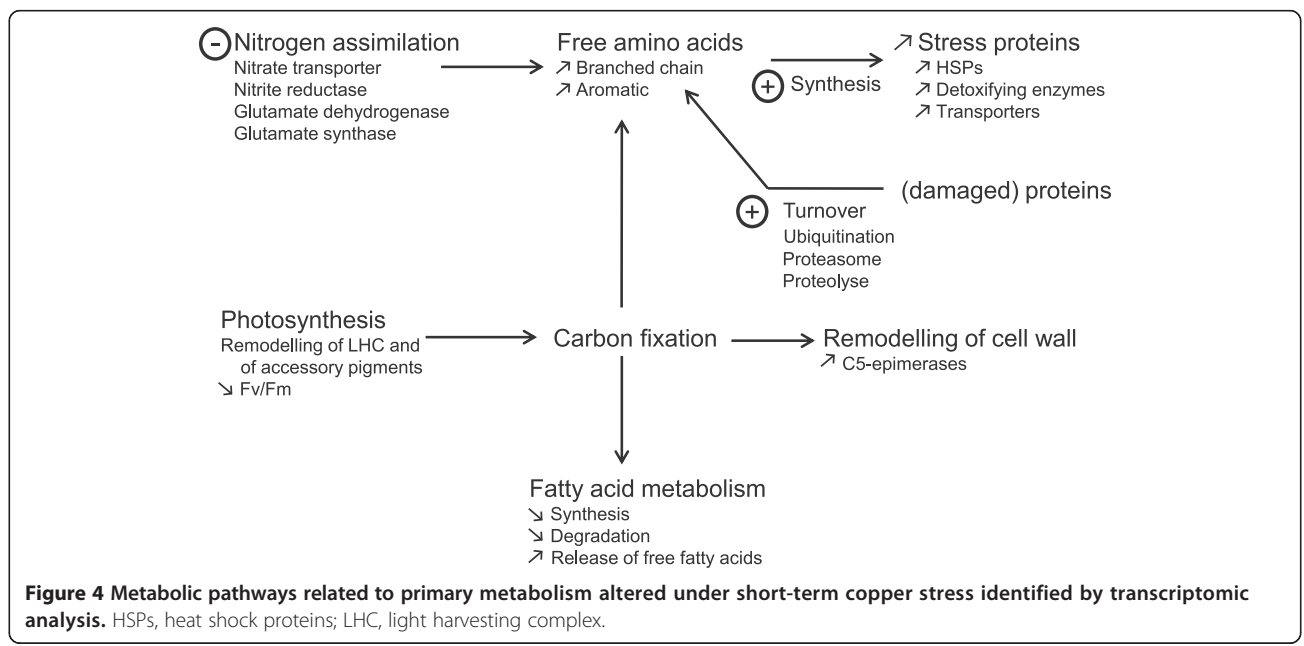

**Figure 4 Metabolic pathways related to primary metabolism altered under short-term copper stress identified by transcriptomic analysis.** HSPs, heat shock proteins; LHC, light harvesting complex.

*Variations in free fatty acid contents and oxylipin synthesis*
Cu stress induced a significant increase in the levels of free fatty acids (FFAs), especially after 8 h of treatment (two-way ANOVA p < 0.05) (Figure 5). In accordance with this analysis, the PLS co-projection of FFA variations discriminated treatments (control - stress) and exposure time (4–8 h) (Additional file 9, panel B).

Regarding FFAs, Cu treatment induced a 3-fold increase of linolenic acid (C18:3), arachidonic acid (C20:4), and eicosapentaenoic acid (C20:5) (two-way ANOVA, p < 0.05; Figure 5). This increase was correlated with the occurrence of octadecanoid and eicosanoid oxygenated derivatives but only after 8 h of stress and in three of the four biological replicates analyzed. Currently, we do not have any explanation why these observations did not hold true for the fourth biological replicate, which much resembled the other replicates of the same condition with respect to all other examined metabolites as well as gene expression profiles. In the three replicates mentioned above, 13-hydroxy-9Z,11E-octadecadienoic acid (13-HODE) and 13S-hydroxy-9Z,11E,15Z-octadecatrienoic acid (13-HOTrE) contents increased under the 8 h stress treatment, with 16- and 3-fold changes respectively, suggesting the occurrence of a 13-lipoxygenase activity (Figure 5). Within the same context, the content of 12-oxophytodienoic acid (12-oxo-PDA) increased 3-fold after 8 h of Cu treatment and the non-enzymatic accumulation of cyclic C18 phytoprostanes $A_1$ was triggered, supporting the occurrence of ROS-mediated lipid peroxidation processes. Finally, Cu stress induced the accumulation of C20:4 derivatives such as oxo-6E,8Z,11Z,14Z-eicosatetraenoic acid (oxo-ETE; 8-fold change after 8 h of treatment), together with the cyclopentenones prostaglandin A2 and J2 (Figure 6).

Finally, it is worth mentioning that among the 392 monoisotopic peaks from ions detected through UPLC-MS analysis, many unknown metabolites were regulated by exposure to Cu (Additional file 8). The analysis of these compounds was beyond the scope of this report, but their identity and function will be the subject of future studies.

## Discussion

Copper is extremely toxic at high concentrations, and induces oxidative stress by altering electron transfer reactions such as photosynthesis and respiration. In our study, we observed a range of different acclimation processes on the molecular level (transcriptomic and metabolite profiling) before the alga exhibited a decrease in photosynthetic yield. Such a decrease in photosynthetic yield results in the alteration of several physiological processes in *E. siliculosus*, such as the formation of heavy metal-substituted chlorophylls [14], reduced carbon fixation and depletion of reducing equivalents. Effectively, since nitrate reduction directly relies on NADH and $FADH_2$ produced by photosynthesis [35], any condition altering photosynthesis (*e.g.* abiotic stress) may directly affect the regulation of N assimilation and the associated primary amino acid metabolism [36] or vice versa. We observed that Cu-stressed *E. siliculosus* down-regulated genes coding for enzymes related to N assimilation. This down-regulation was not related to a decrease of the pool of free primary amino acids. Moreover, an increase in the content of aromatic amino acids was recorded together with an induction of genes encoding proteins involved in the autophagy process, in particular those related to ubiquitination such as putative E3 ubiquitin ligases, and proteins involved in the proteasome complexes. These observations are similar to previous results published in *E. siliculosus* on response to other stress conditions [25], and suggest that alteration of N assimilation and carbon fixation is compensated by degradation of damaged proteins and recycling of amino acids. Similar processes have been identified in land plants during stress response or in diatoms under nitrogen starvation [37,38]. Concomitantly to the down-regulation of genes involved in N assimilation under copper stress, we observed an increase of free fatty acids. Similar links between amino- and fatty acid metabolism have previously been observed in other algae, such as *Chlamydomonas reinhardtii* [39]. Altogether, these data suggest that a substantial part of the response of *E. siliculosus* to short term copper stress consists in balancing primary metabolic processes. Although we can currently only speculate about the exact physiological benefits of each of the observed adjustments, two important functions may be (1) the compensation for stress-induced changes in photosynthesis and (2) the reduction of the energetic budget for nitrogen assimilation. These changes in primary metabolic processes are undoubtedly tightly tied to the specific stress responses and signaling mechanisms discussed below.

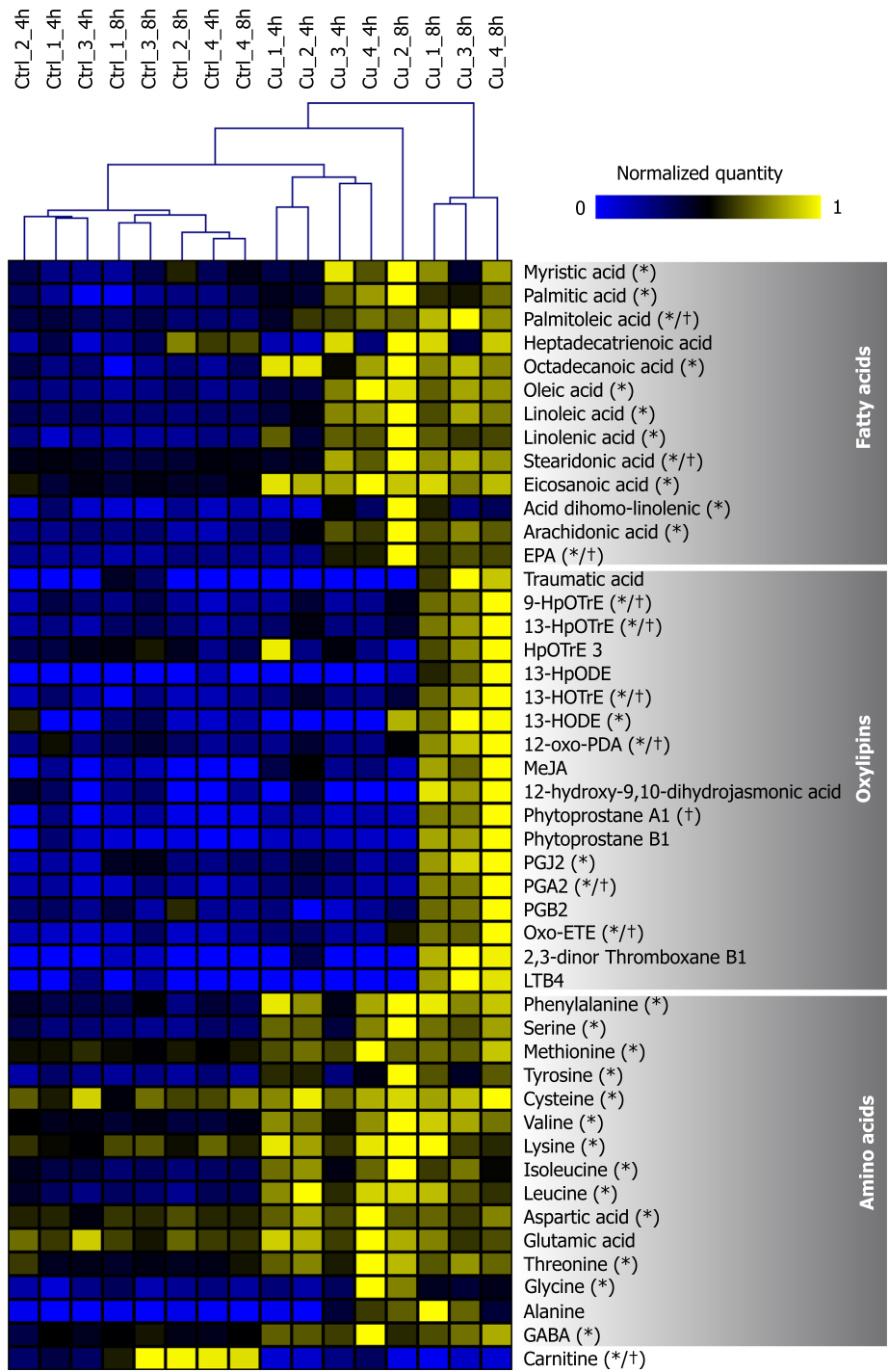

**Figure 5 Heat map of compounds identified by UPLC-MS and GC-MS in copper stress and control conditions.** Samples were arranged according to a hierarchical clustering analysis (Euclidean distance), and the 47 compounds identified were grouped manually. Concentrations of each metabolite were normalized to a maximum of 1 (see Methods section). "*" and "†" indicate significant results (FDR < 5%) in the two-way ANOVA for stress and the interaction term "stress* time", respectively. 9-HpOTrE, 9-hydroperoxy-10,12,15-octadecatrienoic acid; 13-HpOTrE, 13S-hydroperoxy-9Z,11E,15Z-octadecatrienoic acid; HpOTrE 3; oxylipin with the same m/z and raw formula as 11- or 15- hydroperoxy-9Z,11E,15Z-octadecatrienoic acid; 13-HpODE, 13-hydroperoxy-9Z,11E-octadecadienoic acid; 13-HOTrE, 13S-hydroxy-9Z,11E,15Z-octadecatrienoic acid; 13-HODE, 13-hydroxy-9Z,11E-octadecadienoic acid; 12-oxo-PDA, 12-oxophytodienoic acid; MeJA, methyl-jasmonate; PGJ2, prostaglandin J2; PGA2, prostaglandin A2; PGB2, prostaglandin B2; oxo-ETE, oxo-6E,8Z,11Z,14Z-eicosatetraenoic acid; LTB4, leukotriene B4; GABA, γ-aminobutyric acid.

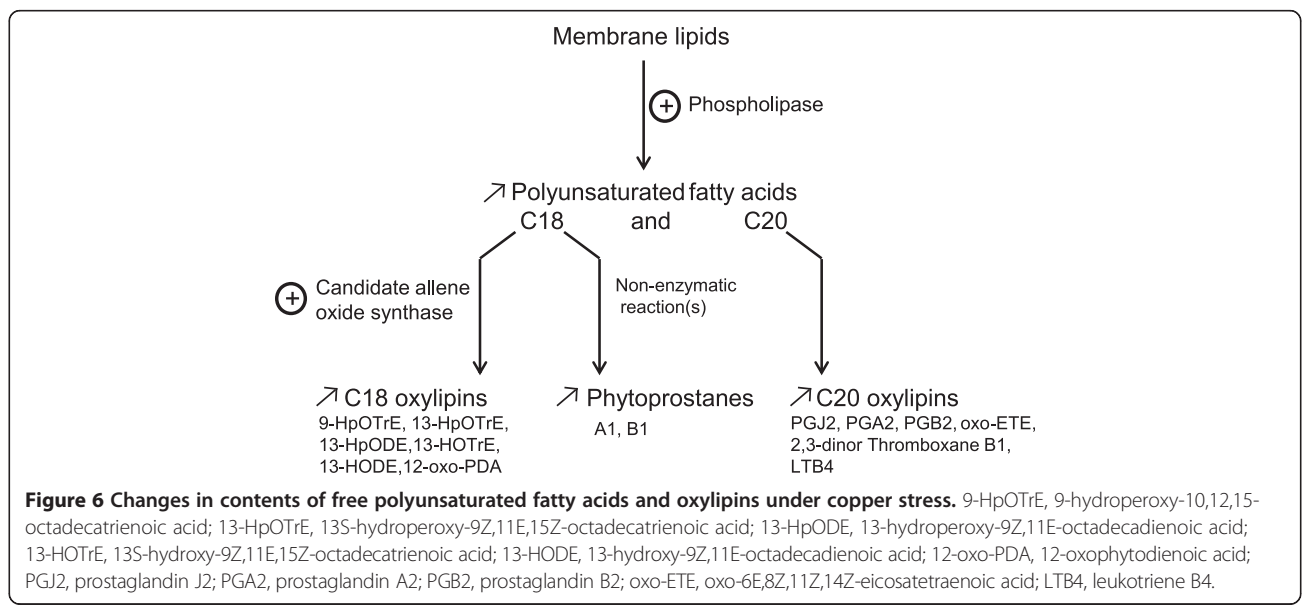

**Figure 6 Changes in contents of free polyunsaturated fatty acids and oxylipins under copper stress.** 9-HpOTrE, 9-hydroperoxy-10,12,15-octadecatrienoic acid; 13-HpOTrE, 13S-hydroperoxy-9Z,11E,15Z-octadecatrienoic acid; 13-HpODE, 13-hydroperoxy-9Z,11E-octadecadienoic acid; 13-HOTrE, 13S-hydroxy-9Z,11E,15Z-octadecatrienoic acid; 13-HODE, 13-hydroxy-9Z,11E-octadecadienoic acid; 12-oxo-PDA, 12-oxophytodienoic acid; PGJ2, prostaglandin J2; PGA2, prostaglandin A2; PGB2, prostaglandin B2; oxo-ETE, oxo-6E,8Z,11Z,14Z-eicosatetraenoic acid; LTB4, leukotriene B4.

Much of the shared Cu and $H_2O_2$ response consisted in genes related to signal transduction and ROS scavenging. This result points out that the Cu-induced ROS formation activates a large set of stress acclimation mechanisms. Apart from the overlapping response, several genes were regulated exclusively under Cu stress, suggesting that this heavy metal may induce the production of ROSs different from those produced under $H_2O_2$ stress, such as $O_2^-$ or $\cdot HO$. In relation to signal transduction, we detected down-regulation of several *E. siliculosus* genes encoding enzymes involved in *myo*-inositol (MI) metabolism, a conserved process in plants and metazoans [10,40]. Recent studies in *Arabidopsis* have demonstrated the importance of MI in oxidative and phytohormone-related stress responses. Using catalase-deficient *A. thaliana* mutant plants, Chaouch and Noctor [41] observed that MI abolished salicylic acid-dependent cell death and pathogen defense responses triggered by peroxisomal $H_2O_2$. As mentioned by Meng *et al.* [42] and Donahue *et al.* [43], a connection has been previously observed between MI synthesis and cell death by studying L-*myo*-inositol 1-phosphate synthase mutants. While our data indicates that a similar link between oxidative stress and the MI pathway may exist in *E. siliculosus*, it remains to be assessed whether the underlying functions or regulatory mechanisms are the same as those found in *A. thaliana*. Besides the direct signaling functions of MI, multi-phosphorylated forms of inositol such as $IP_5$ or $IP_6$ are key cofactors of the plant auxin ($SCF^{TIR1}$) and jasmonate ($SCF^{COI1}$) ubiquitin ligase receptor complexes, therefore playing fundamental roles in phytohormone perception [44]. Methyl-jasmonate (MeJA) was detected in our experiments, but its contents did not change significantly ($p > 0.05$) under the tested conditions. Previous studies have shown that exogenous treatments with MeJA activate defensive mechanisms in several brown algae [45-47]. The synthetic pathway of MeJA has not yet been functionally characterized in brown algae and, moreover, the *E. siliculosus* genome contains neither homologs of land plant JA receptors nor components of the core transcriptional regulatory complex such as MYCs, JAZs, NINJA and COI1 [24,48]. This suggests that this molecule may not exert the same functions in brown algae as in land plants or that this compound has evolved to serve similar functions but using different regulatory systems. However, 8 h of Cu-stress triggered the accumulation of C18 plant-like cyclic oxylipins such as 12-OPDA, $PPA_1$ and $PPA_2$, together with C20:4 cyclic prostaglandins such as $PGA_2$ and $PGJ_2$ in *E. siliculosus*, and this is in agreement with previous observations made in the kelp *L. digitata* [22]. In other organisms these compounds are derived from enzymatic (PGA2, PGJ2, and 12-OPDA) or non-enzymatic (PPA1) processes. All of these fatty acids derivatives contain α,β-unsaturated carbonyl groups and belong to the group of reactive electrophilic species [49]. In land plants, previous studies have demonstrated that 12-OPDA and phytoprostanes play important roles in environmental stress and pathogenesis, acting independently of MeJA, by activating general detoxification and stress responses including HSPs and ABC transporters [50,51]. In this sense, one of the most highly up-regulated genes under copper treatment featuring some sequence similarity with members of the cytochrome P450 CYP74 family is an interesting candidate for AOS activity in *E. siliculosus*, but further experiments will be required to determine the function of this protein with certainty. In our study, we also observed changes in the expression of genes similar to those regulated by RES in land plants. It is therefore reasonable to speculate about (partially) conserved physiological roles of these oxylipins in *E. siliculosus*. Detailed transcriptomic experiments using each of these compounds as elicitor must be carried out to test these hypotheses. However, the increase of octadecanoid and eicosanoid oxygenated fatty acid derivatives suggest that brown algae may use plant-like octadecanoid signals and animal-like eicosanoid oxylipins, which are absent in vascular plants, to regulate protective mechanisms or stress responses.

Metal transport is a key process in the interaction of algae with their environments [52], and Cu treatment induced several potential membrane transport mechanisms in *E. siliculosus*. Of special interest was the up-regulation of a putative ortholog of *A. thaliana* $Cu^{2+}$ $P_{1B}$-type ATPases. Such proteins play an important role in Cu compartmentalization and excretion in plant roots, and corresponding mutants in *A. thaliana* are Cu-hypersensitive [12]. In the same vein, five ABC transporters were up-regulated in response to Cu stress, two of which belong to the B and C sub-families known to participate in the transport of xenobiotics, and of glutathione or phytochelatin conjugates in land plants and animals [31,32]. Additionally, members of the sub-family C, such as the human MRP1, have been shown to act as transporter of eicosanoids such as leukotrienes [31], suggesting a possible link between the concomitant induction of an MRP homolog in *E. siliculosus* and of C20 oxylipin production in our study. Furthermore, several putative transporters of the sub-family A are regulated under Cu stress in *E. siliculosus* and might be involved in lipid trafficking processes in regard to the function of these proteins in land plants and mammals [53,54]. Detoxification transporters of the MATE family and GRX/malate fusion proteins also seem to play a role in Cu-detoxification in *E. siliculosus*. In rice, MATEs are involved in aluminium (Al) resistance by secreting Al-citrate conjugates [55]. Assuming similar functions in *E. siliculosus*, our observation would suggest that the up-regulation of genes encoding these transporters is related to the secretion of Cu-conjugates, and thus that some processes involved in Cu detoxification are conserved across very distant lineages. In contrast, among the mechanisms that seem to be specific to brown algae, one is based on halide metabolism. The *E. siliculosus* genome contains only one gene coding for a putative vBPO, and this gene was up-regulated under Cu stress. The accumulation of this vBPO protein and an increase in the corresponding enzyme activity in both a copper-tolerant and a copper-sensitive strain of *E. siliculosus* (*i.e.* the strain used in this study) in response to chronic Cu stress conditions have previously been reported [21], supporting the potential role of this enzyme in specific brown algal ROS detoxification processes.

## Conclusions

This study provides an overview of copper stress acclimation in *E. siliculosus*, and highlights a number of processes that seem to be conserved with metazoans or land plants. Of interest for future brown algal research is the potential cross-talk between ROS, MI, and oxylipin signaling. Considering that two thirds of the regulated genes identified through this work encode proteins of unknown function, many additional components of the stress response remain to be discovered. The number of available medium to high throughput datasets related to stress response in brown algae is still limited [56], and the analysis of such results paves the way to decipher some of the molecular mechanisms leading to acclimation of these organisms to their frequently changing environment. They will be complemented by follow-up studies benefiting from the development of new techniques and protocols for brown algae, such as reverse genetics (transformation, RNAi, and tilling) or the assessment of nucleic acid-protein and protein-protein interactions.

## Methods

### Cultivation of plant material and experimental set-up

*E. siliculosus* (Ectocarpales, Phaeophyceae) unialgal strain 32 (CCAP accession 1310/4) was cultivated in Provasoli-enriched natural seawater at 14°C using a 14/10 h photoperiod and a photosynthetically active radiation (PAR) intensity of 40 $\mu$mol m$^{-2}$ s$^{-1}$ provided by Philips daylight fluorescent tubes, in 10 L Nalgene flasks under constant aeration. Experiments were carried out with 0.45 $\mu$m filtered natural seawater (FSW) that was collected offshore at Roscoff (+48°46′40,-3°56′15), a site with no direct chemical influence from the shore and presenting dissolved copper concentrations ranging from 5 to 10 nM (Riso *et al.* personal communication). FSW was then treated to avoid potentially chelating organic matter by adding 0.2 g L$^{-1}$ activated charcoal (Merck, Germany) overnight. Treated water was filtered at 0.45 $\mu$m to remove charcoal. Copper stress was triggered by transferring the algae to fresh seawater free of organic matter and enriched with Cu(II) as CuCl$_2$ (Merck, Germany) at nominal final concentrations of 250 or 500 $\mu$g L$^{-1}$ (i.e. 1.8 $\mu$M and 3.7 $\mu$M), in 1 L glass flasks washed overnight with 1% HCl to limit Cu adsorption. Neither nutrients, nor EDTA, were added during the experiments. After 4 and 8 h, replicate cultures were harvested by gentle filtration, algal material was briefly dried using a paper towel, and immediately frozen in liquid nitrogen. For metabolomic profiling, four replicate cultures (250 $\mu$g L$^{-1}$ of CuCl$_2$, as well as corresponding controls) were harvested at each time point. Transcriptomic profiling was carried out in triplicate.

### Measurement of PSII maximum quantum yield

To monitor the stress intensity caused by Cu treatments in algal tissues, maximum quantum yield of PSII (dark adapted) was measured in one-hour intervals as $F_v/F_m$ using a Walz Phyto-PAM (Waltz, Germany) as previously described [21]. Since photosynthesis is stress-sensitive, the quantum yield decreased under sub-optimal (stressful) conditions. Statistically significant differences in Fv/Fm values between treatments were determined by a z-adjusted Mann–Whitney U test performed using Statistica 5.1 (Statsoft, USA; p < 0.05).

### Double strand cDNA sample preparation and microarray hybridization

RNA was extracted from approximately 100 mg (wet weight) of tissue following Apt *et al.* [57] with modifications as described by Le Bail *et al.* [58], using a CTAB-based extraction buffer and subsequent phenol-chloroform purification, LiCl-precipitation, and DNAse (Turbo DNAse, Ambion, Austin, USA) steps. RNA quality and quantity was then verified on a 1% agarose gel stained with ethidium bromide and a NanoDrop ND-1000 spectrophotometer, respectively. Double strand cDNA synthesis and amplification were carried out with the SMART cDNA synthesis kit (Clontech, Mountain View, USA) and the M-MuLV reverse transcriptase (Finnzymes, Espoo, Finland) starting from 100 ng of total RNA, according to a protocol validated by Dittami *et al.* [25]. The optimal number of amplification cycles was determined by semi-quantitative PCR, and ranged from 20 to 25 for our samples. The PCR products were purified by first vortexing with one volume of phenol:chloroform:isoamyl alcohol (25:24:1), then by precipitating the aqueous phase with 0.5 volumes of 7.5 M NH$_4$OH, 6 $\mu$g of nuclease-free Glycogen (Ambion, Austin, USA), and 2.4 volumes of ethanol. After centrifugation at RT (20 min at 14,000 g), the pellet was washed with 70% EtOH, centrifuged (10 min at 14,000 g and RT) and resuspended in 14 $\mu$L of H$_2$O. To finish, cDNAs were run on an agarose gel and quantified with a NanoDrop ND-1000 spectrophotometer, to ensure that they met the requirements for hybridization by Roche NimbleGen (Madison, WI, USA) (concentration > 250 $\mu$g L$^{-1}$, A$_{260/280}$ ≥ 1.7, A$_{260/230}$ ≥ 1.5, median size ≈ 400 bp). Samples were Cy3-labeled and hybridized to the *E. siliculosus* EST-based Roche NimbleGen 4-plex expression array [ArrayExpress:A-MEXP-1445], described in [25]. It is based on a set of 90,637 expressed sequence tags, corresponding to 8,165 contigs and 8,874 singletons, although in some cases, several contigs/singletons belonging to the same genomic locus had not been assembled. In total, the 17,039 contigs/singletons cover about 10,600 of the 16,256 predicted unique genes in the *E. siliculosus* genome [59]. Each of the contigs/singletons is represented by four unique 60-mer probes on the array.

### Validation of microarray results using quantitative PCR

To assess the reliability of fold-change estimations in mRNA abundance by microarray, sixteen genes that exhibited diverse expression patterns in the microarray analysis were analyzed by real time quantitative PCR (RT-qPCR): eight up-regulated, five down-regulated, and three without significant changes (Additional file 10). RT-qPCR was performed as previously described [58], using the three experimental replicates that were also employed in the microarray experiment. Briefly, SYBR green assays were used to determine cDNA copy numbers based on gDNA standard curves ranging from 37 to 48671 copies of the *E. siliculosus* genome. All primer pairs used are given in Additional file 10. Amplification efficiencies were determined for each primer pair and ranged from 90–110%. Furthermore, melting curve analyses were run to confirm that only one product was amplified, and the mN primer pair targeting an intron was used to verify the absence of any genomic DNA [58]. The geometric mean of two reference genes, EF1 (translation elongation factor 1 alpha) and UBCE (ubiquitin conjugating enzyme) was then used to normalize gene expression. Most tested genes had similar expression profiles in both the microarray and the RT-qPCR experiment. We observed an overall correlation $r^2$ of 0.82 between the experiments, indicating good reproducibility (Additional file 10). Different expression patterns between microarray and qPCR were obtained only for two genes coding for a putative zinc/iron permease (ZnFePer) and a tyrosinase-like protein (TYR).

## Statistical analysis of microarray experiments
Normalization and probe averaging of raw microarray data were carried out by Roche Nimblegen using quantile normalization [60], and the robust multichip average algorithm [61]. Normalized microarray signals were then log2-transformed and analysed in a two-way ANOVA with time (4 h, 8 h) and stress (0 or 250 $\mu g$ L$^{-1}$ Cu) as factors using TIGR MeV 4.4 [62]. In this analysis, the false discovery rate (FDR) was limited to 5% according to Benjamini and Hochberg [63]. In parallel, a meta-analysis was carried out to compare our data to a previous dataset available for other stressors [25]. Transcripts previously found to be significantly regulated (p < 0.05, fold-change compared to control > 2) after either 6 h [25] or 4 h or 8 h (this study) were taken into account, but to reduce the impact of false negatives on this analysis no FDR correction was applied here. The results of the meta-analysis are therefore useful to determine which conditions a gene was regulated in, but not to generate lists of significant genes. Hierarchical clustering of data both from Cu stress- and previously published oxidative and salt stress experiments [25] was carried out with TigrMEV version 4.8 using the HCL-support tree function (100 bootstrap replicates). For this analysis, log2-ratios (stress/control) were calculated for each replicate, and Euclidian distance was used as distance matrix.

## Automatic annotation of contigs/singletons
For the purpose of generating automatic genome annotations, the contigs/singletons used for the microarray design [25] were replaced by the full coding sequence using data from the *E. siliculosus* genome project [24]. To this end, all contigs/singletons were blasted against the complete data set of predicted mRNAs (https://bioinformatics.psb.ugent.be/gdb/ectocarpus/Archive/Ectsi_mRNA_Aug2011.tfa.gz). Different thresholds were tested for these blast searches, with e-value cutoffs from 1e-20 to 1e-5, and identity cutoffs from 80% to 99%. Only identity-cutoffs > 95% and e-value cutoffs < 1e-15 resulted in a reduction of the number of sequences with a blast hit. After manual examination of the blast results in question, an identity-cutoff of 90% and an e-value cutoff of 1e-10 were chosen for further analyses. If a blast hit was found to match these criteria, the original sequence was replaced by the complete mRNA sequence for annotation purposes. These results were also used to determine the corresponding genomic loci throughout the manuscript. If no corresponding genomic locus was found using this approach, previously generated correspondences available [64] were used, as these allowed for longer UTRs than those predicted in the genome. All sequences presenting differential expression were automatically annotated and classified following a procedure previously described [25]. Briefly, annotation was carried out with KEGG orthology (KO) numbers using the KO-Based Annotation System KOBAS [65] and with Gene Ontology (GO) terms [66] using the GO Term Prediction and Evaluation Tool GOPET [67].

## Phylogenetic analysis and protein fold prediction
Candidate ABC transporters were identified in the *E. siliculosus* genome based on manual annotations [24] or the presence of the "ABC_tran" PFAM motif, and classified according to Dean *et al.* [31]. To this end, protein sequences of classified human [31] and *A. thaliana* [32] ABC transporters, as well as the candidate *E. siliculosus* sequences, were aligned to the "ABC_tran" motive with Hmmer3 [68]. The resulting alignment formed the basis for maximum likelihood phylogenetic reconstruction using PhyML 3.0 [69]. Analyses were carried out with the LG substitution model [70], a discrete gamma model, and an estimated proportion of invariable sites of 0.006, as these parameters were found to best describe the data using ProtTest 2.4 [70]. An approximate likelihood ratio test was used to evaluate the robustness of the best tree and results were plotted using MEGA 5.1 [71]. Protein fold prediction was carried out with the Phyre2 automatic fold recognition server [72] using the intensive modelling mode [73].

## Metabolite profiling and data analysis
After harvesting, samples were freeze-dried and ground in a mortar. For each sample, 250 ng of 12-OH-lauric acid was added as an internal standard. Extraction was carried out from 300 mg (fresh weight) of algal material during 1 h at 4°C with 1 ml MeOH:H$_2$O (8:2) Samples were then centrifuged at 1,500 g for 15 min at 4°C. Aliquots of the supernatants were then used for metabolomic analysis by ultra-high pressure liquid chromatography coupled to mass spectrometry analysis (UPLC-MS), and for gas chromatography coupled to mass spectrometry (GC-MS) profiling.

UPLC was performed using an RSLC Ultimate 3000 from Dionex equipped with a quaternary pump and autosampler. Separations were achieved using an Acclaim RSLC 120 C18 $1.9\,\mu m$ ($2.1 \times 100$ mm) column (Dionex; Thermo Fisher Scientific, Courtaboeuf, France) maintained at 20°C using $5\,\mu L$ injection volume and a flow-rate of $250\,\mu L$ min$^{-1}$. Mobile phase A was composed of 0.1% acetic acid in $H_2O$, and mobile phase B was 0.1% acetic acid in acetonitrile. The gradient consisted of an initial hold at 20% mobile phase B for 2 min, followed by a linear gradient to 100% B in 8 min and a hold for 14 min, followed by re-equilibration for 6 min at 20% B, for a total run time of 30 min. Mass spectrometry was performed on a Thermo Scientific LTQ-Orbitrap Discovery™ mass spectrometer. Scans were collected in both positive and negative ESI mode over a range of $m/z$ 50–1000. Ionization parameters were set as follows: sheath gas 5 psi, auxiliary gas 5 (arbitrary units), sweep gas 0 (arbitrary units), spray voltage 2.7 kV, capillary temperature 300°C, capillary voltage 60 V, tube lens voltage 127 V and heater temperature 300°C. The Xcalibur 2.1 software was used for instrument control and data acquisition. Raw files were converted to the mzXML format using MassMatrix File Conversion Tools (Version 3.9, April 2011). Data were processed by XCMS [74] running under R or on the online version, using the parameters listed in Additional file 11, and further annotated by CAMERA [75]. This approach allowed detecting 392 monoisotopic peaks, among which 31 free fatty acids and oxylipins were further identified by comparison with standards and mass spectral databases Metlin, HMDB and MassBank.

GC-MS was used for profiling of amino acids. To this end, $100\,\mu L$ aliquots of the MeOH:$H_2O$ (8:2) fraction were evaporated under a stream of nitrogen and resuspended in $60\,\mu L$ of 20 g L$^{-1}$ methoxyamine-hydrochloride (Sigma-Aldrich, Saint-Quentin Fallavier, France) in pyridine before incubation at 60°C for 1 h. After addition of $50\,\mu L$ of N,O-bis(trimethylsisyl)trifluoroacetamide, samples were incubated 1 h at 40°C and then at room temperature overnight before injection. Derivatized metabolites were analyzed by GC-MS using an Agilent GC 6890+ coupled to a 5975 MS Detector (Agilent, Les Ulis, France) and equipped with a DB-5MS column (30 m × 0.25 mm I.D. × $0.25\,\mu m$ film thickness; J&W Scientific, Agilent) in the EI mode at 70 eV. The temperature gradient used was 60°C for 5 min, 60–120°C at 30°C min$^{-1}$, 120–290°C at 4°C min$^{-1}$, and 290°C for 10 min. Sixteen amino acids were identified by comparison with standards.

By combining the data obtained in both positive and negative ion mode by UPLC-MS and by GC-MS, 47 metabolites corresponding to fatty acids, oxylipins, and amino acids could be reliably identified. Multivariate statistical analyses of metabolite data were carried out using SIMCA-P (12.0.1, Umetrics, Umeå, Sweden). Data were log10-transformed and normalized using Pareto scaling. Partial least squares discriminant analysis (PLS-DA) was carried out on the 392 monoisotopic peaks detected by the UPLC-MS analysis, and also independently on free fatty acids and on the amino acids. Hierarchical clustering analysis of the 392 peaks was carried out with TigrMeV as described for gene expression data, except that metabolite quantities were normalized by dividing all values by the highest value obtained in any of the conditions tested (all normalized values ranged from 0 to 1) instead of calculating log2-ratios. A similar analysis was also performed for the 47 identified metabolites, and their relative quantities were further analysed using a two-way ANOVA in analogy to the gene expression data (see above).

## Availability of supporting data

## Additional files

**Additional file 1: List of the 627 contigs/singletons found to be differentially regulated under Cu treatments at an FDR of 5%.** Sequences were manually grouped into functional categories. Values highlighted in yellow correspond to genes up-regulated, while blue background indicates genes down-regulated.

**Additional file 2: Venn diagram representing the number of significantly up-regulated (a) and down-regulated (b) contigs/singletons under copper (Cu), oxidative, hypersaline (Hyper), and hyposaline (Hypo) stress conditions (p < 0.05).**

**Additional file 3: List of contigs/singletons identified as differentially regulated (p < 0.05; fold-change > 2; no FDR correction) under different abiotic stress conditions tested through the meta-analysis.** Protein definitions were retrieved from the *E. siliculosus* genome database in March 2013.

**Additional file 4: Hierarchical clustering of gene expression data obtained by microarray in *E. siliculosus* submitted to different abiotic stress conditions.** Hypo_1-4 = hyposaline stress (salinity of 4 ppt, 6 h), Hyper_1-4 = hypersaline stress (salinity of 96 ppt, 6 h), Oxi_1-4 = oxidative stress (10 mM $H_2O_2$, 6 h; see Dittami *et al.* [25] for experimental details for these stressors), and Cu_1_4h - Cu_3_8h (250 μg L$^{-1}$ copper stress, 4–8 h; this study). The heat map shows log2-ratios of gene expression in stress compared to control conditions. Gene trees and sample trees were obtained by hierarchical clustering and "Euclidean distance" as metric.

**Additional file 5: (A) ClustalW multiple alignment of the CYP74 family IHCD domains from Esi0060_0078 and homologs.** The conserved catalytic residues FXXXFXSXX were highlighted in red for the CYP74A conserved phenylalanine residues and blue for the AOS specific serine residue. (B) Ribbon diagram of a predicted structural model of Esi0060_0078 prepared with the Phyre2 software [73].

**Additional file 6: Summary of features observed in the putative *E. siliculosus* Cu induced heavy metal P$_{1B}$-ATPase (Esi0023_0054).** This sequence contains conserved heavy metal-associated (HMA), E1-E2 ATPase superfamily and haloacid dehalogenase (HAD) domains. The diagram was obtained by amino acid sequence analysis based on the Conserved Domains and Protein Classification Database.

**Additional file 7: Results of partial least squares discriminant analysis (PLS-DA) carried out for the compounds detected in the Cu-stress experiment.** The plot represents the variations of 392 monoisotopic peaks quantified by UPLC-MS in positive ion mode in algal samples harvested after 4 h and 8 h of Cu-stress (red and yellow spots, respectively), as well as the corresponding controls (green and blue spots).

**Additional file 8: Hierarchical clustering of 392 monoisotopic peaks quantified by UPLC-MS in positive and negative ion mode in algal samples under copper stress and control conditions.** Concentrations of each metabolite were normalized to a maximum of 1 (see Methods) and clustering was carried out with the Euclidean distance matrix.

**Additional file 9: Results of PLS-DA representing the variations of (A) amino acid (GC-MS analysis) and (B) fatty acid contents (UPLC-MS analysis) between control and Cu treated samples.**

**Additional file 10: Primers used for qPCR experiments, and comparison of results obtained by microarray and qPCR analysis for selected genes.** In the included graph, each point represents the Log2-ratio of the mRNA abundance determined by microarray in copper stress and in control conditions for a specific gene after 4 h or 8 h of Cu treatment (mean of three biological replicates) plotted against the same ratio determined by qPCR.

**Additional file 11: Parameters used for XCMS analysis.**

**Competing interests**
The authors declare that they have no competing interests.

**Authors' contributions**
AR, SD, SG, JC, CB, PP, and TT conceived the study; AR and SG carried out the experiments; AR, SD, SG, and TT analyzed the data and wrote the manuscript; all authors corrected and approved the final manuscript.

**Acknowledgements**
We are especially grateful to Laurence Dartevelle for valuable help with Ectocarpus siliculosus cultivation. This work has been partially funded by Marine Genomics Europe NoE 3 (EU contract n° GOCE-CT-2004-505403), and the French Embassy and the CONICYT of Chile through PhD fellowship to A.R. This work was also supported by the Laboratoire International Associé 'Dispersal and Adaptation of Marine Species' (LIA DIAMS) PUC, Chile, and CNRS-UPMC, France. S.G. benefited from the support of the French Government via the National Research Agency through the investment expenditure program IDEALG (ANR-10-BTBR-04). Additional support came from FONDAP 1501–0001 (Program 7) to J.A.C. and A.R.

**Author details**
¹UPMC Univ Paris 06, UMR 8227, Integrative Biology of Marine Models, Station Biologique de Roscoff, Sorbonne Universités, CS 90074, F-29688 Roscoff cedex, France. ²CNRS, UMR 8227, Integrative Biology of Marine Models, Station Biologique de Roscoff, CS 90074, F-29688 Roscoff cedex, France. ³Departamento de Ecología, Center of Applied Ecology & Sustainability, Facultad de Ciencias Biológicas, Pontificia Universidad Católica de Chile, Santiago, Chile. ⁴Plate-forme MetaboMER, CNRS & UPMC, FR2424, Station Biologique, 29680 Roscoff, France. ⁵Present addresses: Department of Plant Systems Biology, VIB and Department of Plant Biotechnology and Bioinformatics, Ghent University, Technologiepark 927, Ghent B-9052, Belgium.